\documentclass{llncs}

\usepackage[latin1]{inputenc}
\usepackage{graphicx}
\usepackage{algorithm, algorithmic}
\usepackage{xspace}
\usepackage{subfigure}

\begin{document}
  \title{Efficient searching and retrieval of documents in PROSA}
  \def\thealg{\textit{\textbf{PROSA}}\xspace}
  
  \author{Vincenza Carchiolo\inst{1}
    \and Michele Malgeri\inst{1} 
    \and Giuseppe Mangioni\inst{1} 
    \and Vincenzo Nicosia \inst{1}}
  
  \institute{ Universit\`a degli Studi di Catania\\ 
    Facolt\`a di Ingegneria \\
    V.le A. Doria 6 \\
    95100 -- Catania}
  
  \maketitle
  \begin{abstract}
Retrieving resources in a distributed environment is more difficult
than finding data in centralised databases. In the last decade P2P
system arise as new and effective distributed architectures for
resource sharing, but searching in such environments could be
difficult and time--consuming. In this paper we discuss efficiency of
resource discovery in \thealg, a self--organising P2P system heavily
inspired by social networks.  All routing choices in \thealg are made
locally, looking only at the relevance of the next peer to
each query.  We show that \thealg is able to effectively answer
queries for rare documents, forwarding them through the most
convenient path to nodes that much probably share matching
resources. This result is heavily related to the small--world
structure that naturally emerges in \thealg. 
\end{abstract}

\section{Introduction}

Organisation of electronical resources and documents is of the most
importance for efficient searching and retrieval. Nowadays the WWW is
a (negative) example of how searching and obtaining informations from
an unstructured knowledge base could really become difficult and
frustrating.  In the case of the World Wide Web, this problem is faced
and partially resolved by centralised searching engines, such as
Google, MSN--Search, Yahoo and so on, which can help users in pruning
away unuseful resources during searches. But searching strategies used
by web indexing engines cannot be easily adopted in a P2P
environment, mainly because nodes of such a distributed system cannot
be compared to web--servers. Each peer shares a small amount of
resources, can join and leave the network many times in a week and
usually searches and retrieve resources belonging to a small number of
different topics.  In the last few years many P2P structures have been
proposed, in order to build a valuable and efficient distributed
environment for resource sharing.

The problem is that existing P2P systems usually ask the user to
choose between efficiency and usability. In fact, while DHT systems
allow fast resource searching \cite{clarke01freenet}
\cite{ratnasamy00scalable} \cite{zhao01tapestry} introducing unnatural
indexing models, unstructured and weakly structured P2P systems
\cite{Hybrid}\cite{GES}\cite{SETS} usually allow users to easily
express queries but have poor performance with respect to bandwith and
time consumption.

In this work we analyse retrieving performance of \thealg (P2P
Resource Organisation by Social Acquaintances), a P2P
system heavily inspired by social networks: joining, searching
resources and building links among peers in \thealg are performed in
a social way. Each peer gains a certain amount of strong links to
peers which share similar resources and also maintains weak links to
far away peers. 

The linking phase is similar to a birth: each peer is given just a
couple of weak links which can be used for query forwarding. Queries
for resources are forwarded through outgoing links to other peers, in
accordance with a defined ``similarity'' between the query and shared
resources. New relationships in real social networks arise because
people have similar interests, culture and knowledge. In a similar
way, new links among peers in \thealg are established when a query is
forwarded and successfull answered, so that peers which share similar
resources finally get connected together. 

In this paper we focus on the ability of \thealg in answering queries
with a sufficient number of results, even if a small amount of
existing documents match them. Matching documents are retrieved in an
efficient way, forwarding queries to a small amount of nodes using
just a few ``right'' links, thanks to small--world structure that
naturally emerges in a \thealg network.

In section \ref{sect:alg} we give a brief formal description of
involved algorithms ; section \ref{sect:info} reports simulation
results, focused on retrieval of rare resources; in section
\ref{sect:energy} the efficiency of the query routing algorithm is
discussed, while section \ref{sect:future} propose guidelines for
future work.

\section{\thealg: a brief description }
\label{sect:alg}

\label{s:prosa}
As stated above, \thealg is a P2P network based on social
relationships.  More formally, we can model \thealg{} as a directed
graph: \begin{equation} \label{equ:prosa} \thealg = (\mathcal{P},
\mathcal{L}, P_r, Label) \end{equation}

$\mathcal{P}$ denotes the set of peers (i.e. vertices), $\mathcal{L}$
is the set of links $l = (s, t)$ (i.e. edges), where $t$ is a neighbour
of $s$.  For link $l = (s, t)$, $s$ is the source peer and $t$ is the
target peer. All links are directed.

In P2P networks the knowledge of a peer is represented by 
resources it shares with other peers.  In \thealg the mapping $P_r:
\mathcal{P} \rightarrow 2^{\mathcal{R}}$, associates peers with
resources. For a given peer $s \in \mathcal{P}$, $P_r(s)$ is the 
set of resources hosted by peer $s$. 
Given a set of resources, we define a function 
$R_c: 2^\mathcal{R} \rightarrow \mathcal{C}$ 
that provides a sort of compact description of all resources.
We also define a function
$P_k: \mathcal{P} \rightarrow \mathcal{C}$, such that, for a given 
peer $s$, $P_k(s)$ is a compact
description of the peer knowledge ($PK$ - {\em Peer Knowledge}).
It can also be obtained combining $P_r$ and $R_c$: $P_k (s) = R_c (P_r (s))$.

Relationships among people in real social networks are usually based
on similarities in interests, culture, hobbies, knowledge and so on
\cite{newman-2001-98}\cite{scott-handbook}\cite{albert-2002-74}.
Usually these kind of links evolve from simple ``acquaintance--links''
to what we called ``semantic--links''.  To implement this behaviour
three types of links have been introduced: {\em Acquaintance--Link}
($AL$), {\em Temporary Semantic--Link} ($TSL$) and {\em Full
Semantic--Link} ($FSL$).  TSLs represent relationships based on a
partial knowledge of a peer. They are usually stronger than $AL$s and
weaker than $FSL$s.

In \thealg, if a given link is a simple $AL$, then the source peer
does not know anything about the target peer.  If the link is a $FSL$,
the source peer is aware of the kind of knowledge owned by the target
peer (i.e. it knows $P_k (t)$, where $ t \in \mathcal{P}$ is the
target peer). Finally, if the link is a $TSL$, the peer does not know
the full $P_k(t)$ of the linked peer; it instead has a {\em Temporary
Peer Knowledge} ($TP_k$) which is based on previously received queries
from the source peer.  Different meanings of links are modelled by
means of a labelling function $Label$: for a given link $l = (s,t) \in
L$, $Label(l)$ is a vector of two elements $[e, w]$: the former is the
link label and the latter is a weight used to model what the source
peer knows about the target peer; this is computed as follows:
\begin{itemize}
	\item if $e = AL \Rightarrow w = \emptyset$ 
	\item if $e = TSL \Rightarrow w = TP_k$
	\item if $e = FSL \Rightarrow w = P_k(t)$ 
\end{itemize}

In the next two sections, we give a brief description of how
\thealg works.  A detailed description of \thealg can be found in
\cite{prosa@cops06}.


\subsection{Peer Joining \thealg}
The case of a node that wants to join an existing network is similar
to the birth of a child. At the beginning of his life a child
``knows'' just a couple of people (his parents). A new peer which
wants to join, just looks for $n$ peers at random and establishes
$AL$s to them.  These links are $AL$s because a new peer doesn't know
anything about its neighbours until he doesn't ask them for
resources. This behaviour is quite easy to understand: when a baby
comes to life he doesn't know anything about his parents and
relatives.  The \thealg peer joining procedure is described by
algorithm \ref{join}.

\begin{algorithm}
\caption{\small {\bf JOIN:} Peer $s$ joining to $\thealg
(\mathcal{P},\mathcal{L}, P_r, Label)$}
\footnotesize
\label{join}
\begin{algorithmic}[1]
	\REQUIRE $\thealg (\mathcal{P}, \mathcal{L}, P_r, Label), Peer\:s$
	\STATE $\mathcal{RP} \leftarrow rnd (P, n)$ \COMMENT {Randomly selects $n$ peers of \thealg}
	\STATE $\mathcal{P} \leftarrow \mathcal{P} \cup s$ \COMMENT {Adds $s$ to set of peers}
	\STATE $\mathcal{L} \leftarrow \mathcal{L} \cup \{(s, t), \forall t \in \mathcal{RP}\}$ 
		   \COMMENT {Links $s$ with the randomly selected peers}
	\STATE $\forall t \in \mathcal{RP} \Rightarrow Label(p, q) \leftarrow [AL, \emptyset]$
			\COMMENT {Sets the added links as $AL$}
\end{algorithmic}
\end{algorithm}

\subsection{\thealg dynamics}
In order to show how does \thealg work, we need to define the structure
of a query message.  Each query message is a quadruple:
\begin{equation}
	Q_M = (qid, q, s, n_r)
\end{equation}

where $qid$ is a unique query identifier to ensure that a peer does
not respond to a query more then once; $q$ is the query, expressed
according to the used knowledge model\footnote{If knowledge is
modelled by Vector Space Model, for example, $q$ is a state vector of
stemmed terms. If knowledge is modelled by ontologies, $q$ is an
ontological query, and so on}; $s \in P$ is the source peer and $n_r$
is the number of required results.
\thealg dynamic behaviour is modelled by algorithm \ref{search} and
is strictly related to queries.  When a user of \thealg asks for
a resource on a peer $s$, the inquired peer $s$ builds up a query $q$
and specify a certain number of results he wants to obtain $n_r$. This
is equivalent to call $ExecQuery (\thealg, s, (qid, q, s, n_r))$.

\begin{algorithm}
\caption{\footnotesize{\bf ExecQuery:} query $q$ originating from peer $s$ executed on peer $cur$}
\label{search}
\begin{algorithmic}[1]
\small
	\REQUIRE $\thealg(\mathcal{P}, \mathcal{L}, P_r, Label),\;cur  \in \mathcal{P}, \;q \in QM$
	\STATE $Result \leftarrow \emptyset $
	\IF {$cur \neq s$}
		\STATE $UpdateLink (\thealg, cur, s, q)$
	\ENDIF
	\STATE $(Result,numRes) \leftarrow ResourcesRelevance (\thealg, q, cur, n_r)$
	\IF {$numRes = 0$}
		\STATE $f \rightarrow SelectNextPeer(\thealg, cur, q)$
		\IF {$f \neq null$}
			\STATE $ExecQuery (\thealg, f, qm)$
		\ENDIF
	\ELSE
		\STATE $SendMessage (s, cur, Result)$
		\STATE $\mathcal{L} \leftarrow \mathcal{L} \cup (s, cur)$
		\STATE $Label (s, cur) \leftarrow [FSL, P_k(cur)]$
		\IF {$numRes < n_r$}
			\STATE \COMMENT {-- Semantic Flooding --}
			\FORALL {$t \in Neighborhood(cur)$}
				\STATE $rel \rightarrow PeerRelevance(P_{k}(t), q)$
				\IF {$rel > Threshold$}
					\STATE $qm \leftarrow (qid, q, s, n_r - numRes)$
					\STATE $ExecQuery (\thealg, t, qm)$
				\ENDIF
			\ENDFOR
		\ENDIF
	\ENDIF
\end{algorithmic}
\end{algorithm}

The first time $ExecQuery$ is called, $cur$ is equal to $s$ and this
avoids the execution of instruction \# $3$.  Following calls of
$ExecQuery$, i.e. when a peer receives a query forwarded by another
peer, use function $UpdateLink$, which updates the link between
current peer $cur$ and the forwarding peer $prev$, if necessary. If
the requesting peer is an unknown peer, a new $TSL$ link to that peer
is added having as weight a Temporary Peer Knowledge($TP_k$) based on
the received query message. Note that a $TP_k$ can be considered as a
``good hint'' for the current peer, in order to gain links to other
remote peers. It is really probable that the query would be finally
answered by some other peer and that the requesting peer will
eventually download some of the resources that matched it. It would be
useful to record a link to that peer, just in case that kind of
resources would be requested in the future by other peers.  If the
requesting peer is a $TSL$ for the peer that receives the query, the
corresponding $TP_k$ is updated. If the requesting peer is a $FSL$, no
updates are necessary.

The relevance of a query with respect to the resources hosted by a
peer is evaluated calling function $ResourcesRelevance$.  Two
possible cases can hold: 
\begin{itemize}
\item
  If none of the hosted resources has a sufficient relevance, the query
  has to be forwarded to another peer $f$, called ``forwarder''.  This
  peer is selected among $s$ neighbours by $SelectForwarder$, using the
  following procedure:
	
	\begin{itemize}
	\item[-]
	Peer $s$ computes the relevance between query $q$ and the weight
	of each links connecting itself to his neighbourhood.
	\item[-]
	It selects the link with the highest relevance, if any, and
	forward the query message to it.
	\item[-]
	If the peer has neither $FSL$s nor $TSL$s, i.e. it has just $AL$s,
	the query message is forwarded to one link at random.
	\end{itemize}
This procedure is described in algorithm \ref{search}, where 
subsequent forwards are performed by means of recursive calls to
$ExecQuery$.
	
\item
If the peer hosts resources with sufficient relevance with
respect to $q$, two sub-cases are possible:

	\begin{itemize}
	\item[-] The peer has sufficient relevant documents to full-fill the
	request. In this case a result message is sent to the requesting
	peer and the query is no more forwarded.
	\item[-] The peer has a certain number of relevant documents, but
	they are not enough to full-fill the request (i.e. they are $<
	n_r$).  In this case a response message is sent to the requester
	peer, specifying the number of matching documents. The message query
	is forwarded to all the links in the neighbourhood whose relevance
	with the query is higher than a given threshold (semantic
	flooding). The number of matched resources is subtracted from the
	number of total requested documents before each forward step.
  \end{itemize}
\end{itemize}

When the requesting peer receives a response message it build a new
FSL to the answering peer and then presents results to the
user. If the user decides to download a certain resource from another
peer, the requesting peer directly contacts the peer owning that
resource asking for download. If download is accepted, the resource is
sent to the requesting peer.

\section{Information Retrieval in PROSA}
\label{sect:info}

Other studies about \thealg \cite{prosa@cops06} \cite{prosa@selfman06}
revealed that it naturally evolves to a small--world network, with a
really high clustering coefficient and a relatively small average path
length between peers.

The main target of this work is to show that \thealg does not only
has desirable topological properties, but also that resource
searching can be massively improved exploiting those characteristics.
The fact that all peers in \thealg are connected by a small number of
hops does not guarantees anything about searching efficiency.  In this
section we show that searching resources in \thealg is really fast and
successfull, mainly because peers that share resources in the same
topic usually results to be strongly connected with similar peers.

\subsection{Two words about simulations}
\label{subsect:twowords}

In order to show that \thealg can be used to efficiently share
resources in a P2P environment, we developed a event-driven functional
simulator written in Python. The knowledge base used for simulations
is composed by scientific articles in the field of math and
phylosophy. Articles about math come from ``Journal of American
Mathematical Society''\cite{ams-jams}, ``Transactions of the American
Mathematical Society''\cite{ams-tran} and ``Proceedings of the
American Mathematical Society''\cite{ams-proc}, for a total amount of
740 articles.  On the other hand, articles in the field of philosophy
come from ``Journal of Social Philosophy''
\cite{blackwell-social-phil}, ``Journal of Political Philosophy''
\cite{blackwell-polit-phil}, ``Philosophical Issues''
\cite{blackwell-phil-issue} and ``Philosophical Perspectives''
\cite{blackwell-phil-persp}, for a total amount of 750 articles.

The simulator uses a Vector Space \cite{866292} knowledge model for
resources. Each document is represented by a state vector which
contains the highest 100 TF--IDF \cite{TFIDF} weights of terms
contained into the document.

Each peer contains, on average, $20\pm 5$ articles in the same
topic. Nodes perform 80\% of queries in the same topic of the hosted
resources and the remaining 20\% in the other topic. We choose to do
so after some studies about queries distribution in a Gnutella P2P
system \cite{Hybrid} and with real social communities in mind, where
the most part of requests for resources are focused on a really small
amount of topics.

\subsection{Number of retrieved documents}
\label{subsect:numdocs}

One of the most relevant quality measure of a resource searching
algorithm is the number of documents retrieved by each query. In this
paragraph we examine results obtained with \thealg, using the query
mechanism described in section \ref{s:prosa}. We also compare \thealg
to other searching strategies, such as random walk and flooding.

\begin{figure}[!htb]
  \centering
  \subfigure[Average \# of retrieved documents per query]{
    \includegraphics[scale=0.22]{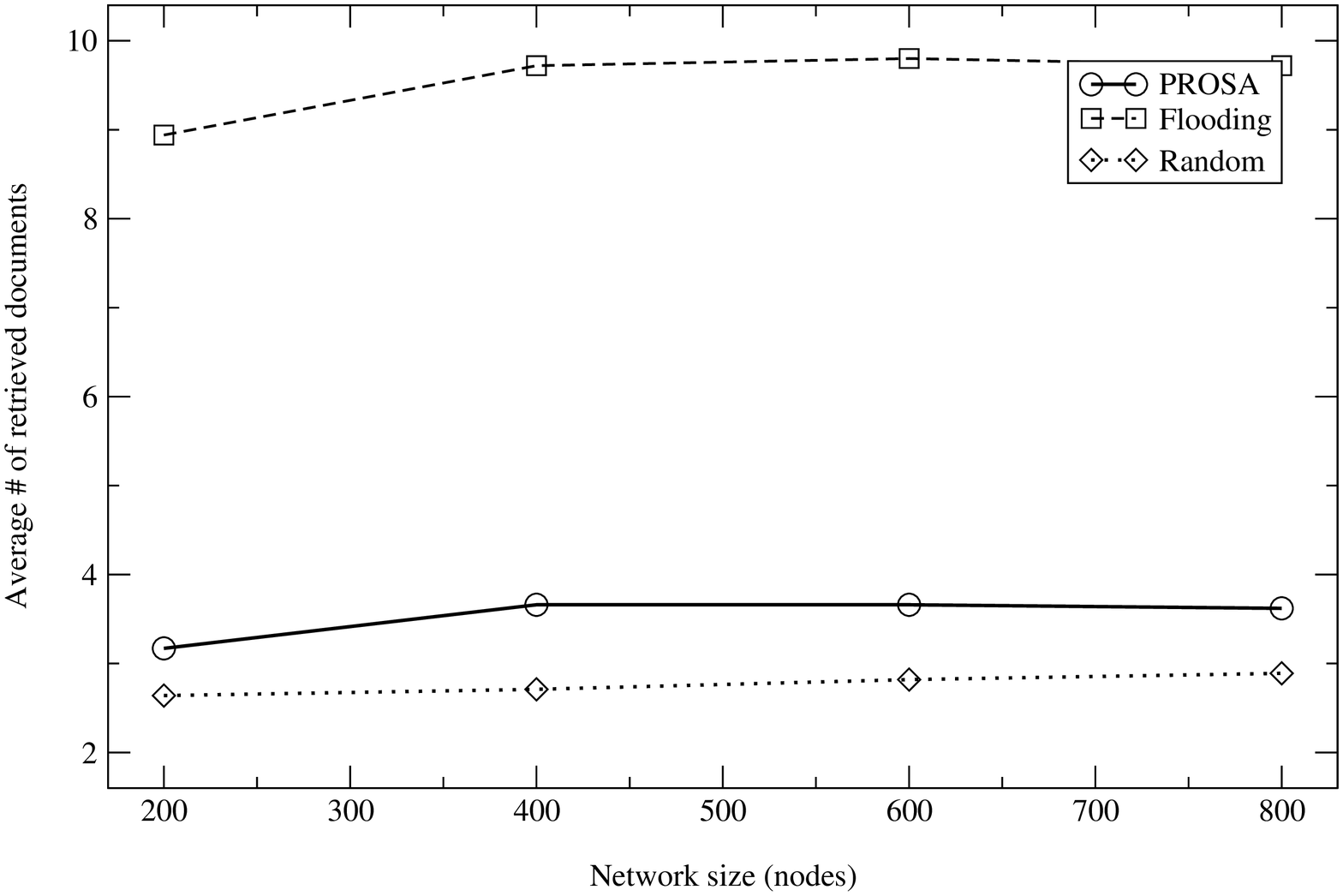}
    \label{fig:docnum}
  }
  \hspace{0.1cm}
  \subfigure[Average \# of retrieved document per successful query]{
    \includegraphics[scale=0.22]{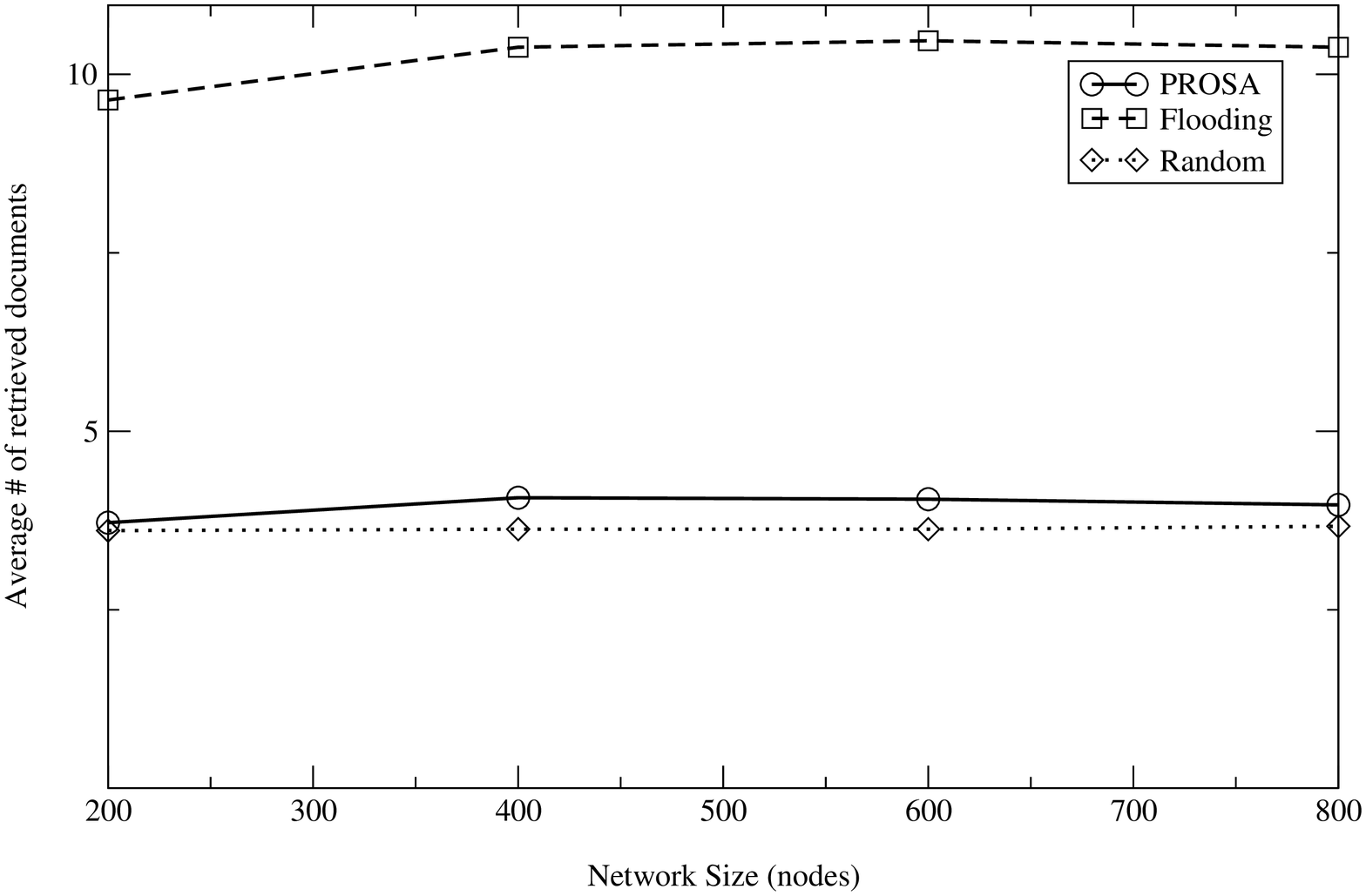}
    \label{fig:numsucc}
  }
  \\
  \subfigure[Percentage of answered queries]{
    \includegraphics[scale=0.22]{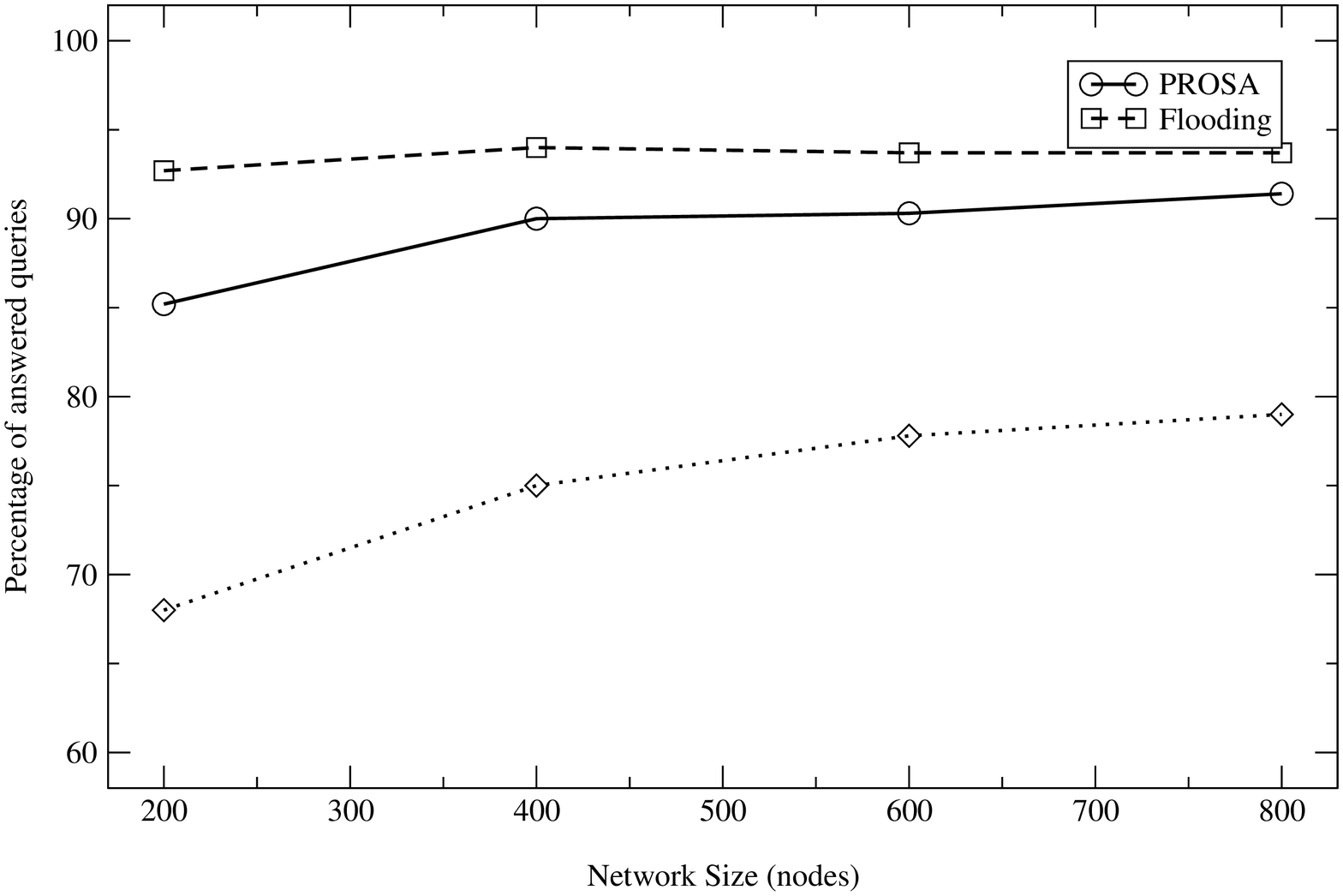}
    \label{fig:percanswer}
  }
  \caption{}
\end{figure}

Figure \ref{fig:docnum} shows a comparison of average number of
retrieved documents in a \thealg network for different number of
nodes, when each node performs 15 queries on average.

\begin{figure}[!htbp]
  \centering
\end{figure}

As showed in figure \ref{fig:docnum}, the best performance is obtained
by flooding, since the average number of retrieved documents per query
is about 10, that is the number of documents required by each query
($n_r$) \footnote{A query is no more forwarded if a sufficient number
of documents has beed retrieved, as explained in \ref{s:prosa}}.
Nevertheless, \thealg is able to retrieve about 4 documents per query,
on average, and this result is still better than that obtained with a
random walk, which usually retrieves only 2.8 documents per query.

This suggests that the query routing algorithm, based on local link
ranking, is really efficient and usually let queries ``flow'' in the
direction of nodes that can probably answer them. We note that \thealg
is able to retrieve a relatively high number of documents also if
compared with a simple flooding. This is a good result, since flloding
is known as beeing the optimal searching strategy: queries are
actually forwarded to all nodes, so all existing and matching
documents are retrieved, until the number of required documents has
not been obtained.

In figure \ref{fig:numsucc} the average number of retrieved documents
per successfull query is reported. The best perormance is once again
obtained by flooding, while \thealg retrieves an average of 4.2
documents for each successfull query over 10 documents
required. Random walk has, once again, the worst performance.

Looking only at the number of retrieved documents could be misleading:
it is not important to have a small amount of queries answered with a
high number of documents. It is desireable having almost all feasible
queries \footnote{A query is feasible if there exist matching
documents to answer it. Otherwise it is considered unfeasible}
answered by a sufficient number of documents.  Figure
\ref{fig:percanswer} shows the percentage of retrieved documents for
\thealg, flooding and random walk, on the same \thealg network with
different network sizes. Note that in every case the average amount of
unfeasible queries is around 6\%.

The highest percentage of answered queries is obtained by flooding the
network, since about 94\% of queries have an answer. This means that
practically all the queries are answered, if we except those that have
no matching documents.  A valuable result is obtained also by \thealg:
84\% to 92\% of all queries are answered, while random walk usually
returns result for less than 80\% of issued queries \footnote{If a
query eventually enters an unconnected component, it cannot be further
forwarded.}.  The percentage of answered queries increases whith
network size, for all searching strategies, because all nodes have an
average number of 20 documents: more nodes means more documents,
i.e. an higher probability of finding matching documents.

\subsection{Query recall}
\label{subsect:recall}

\begin{figure}[!htb]
  \centering
  \subfigure[Query Recall distribution]{
  \includegraphics[scale=0.22]{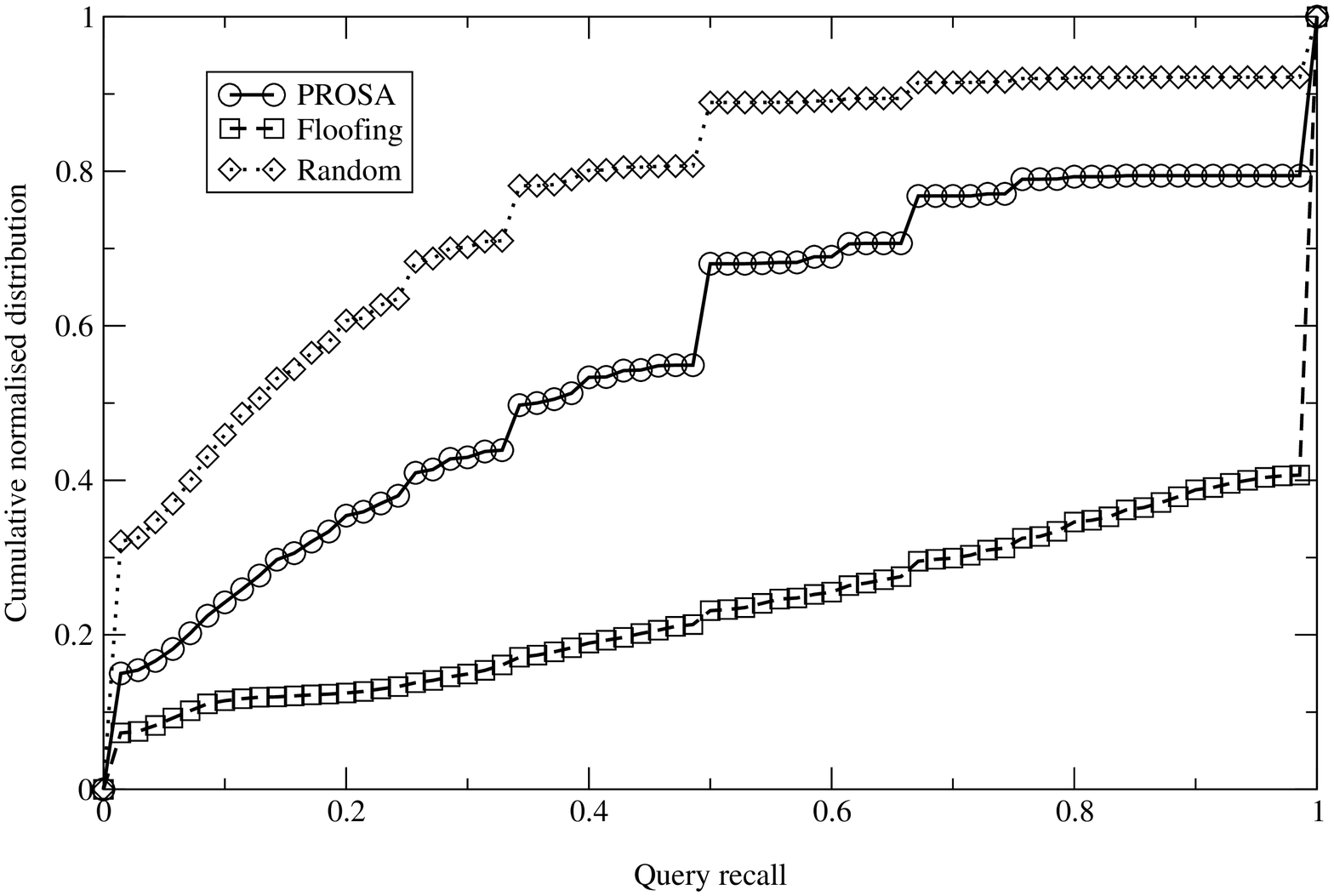}
  \label{fig:recall}
  }
  \hspace{0.1cm}
  \subfigure[Recall distribution for rare queries]{
    \includegraphics[scale=0.22]{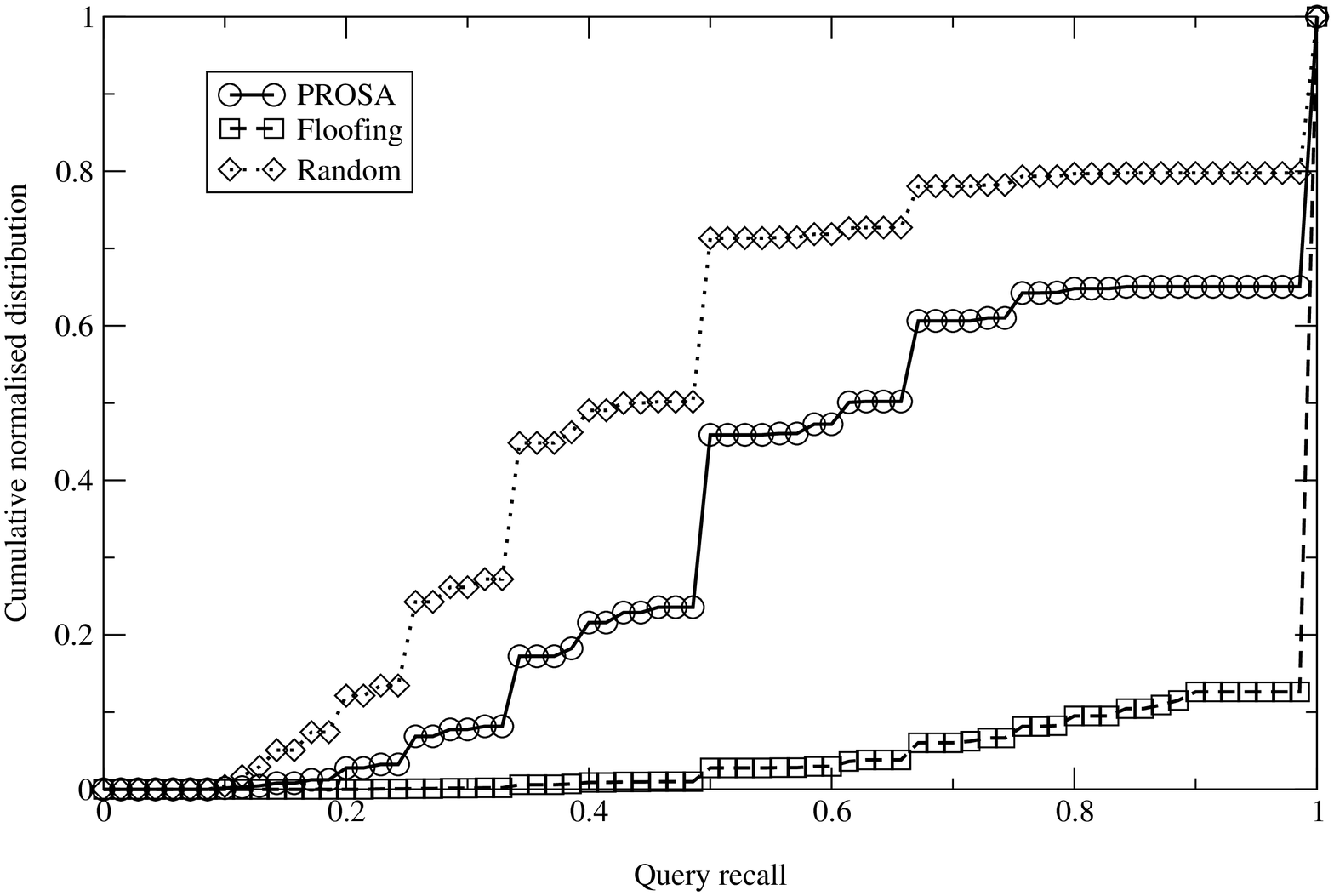}
    \label{fig:rare}
  }\\
    \subfigure[Recall distribution for common queries]{
    \includegraphics[scale=0.22]{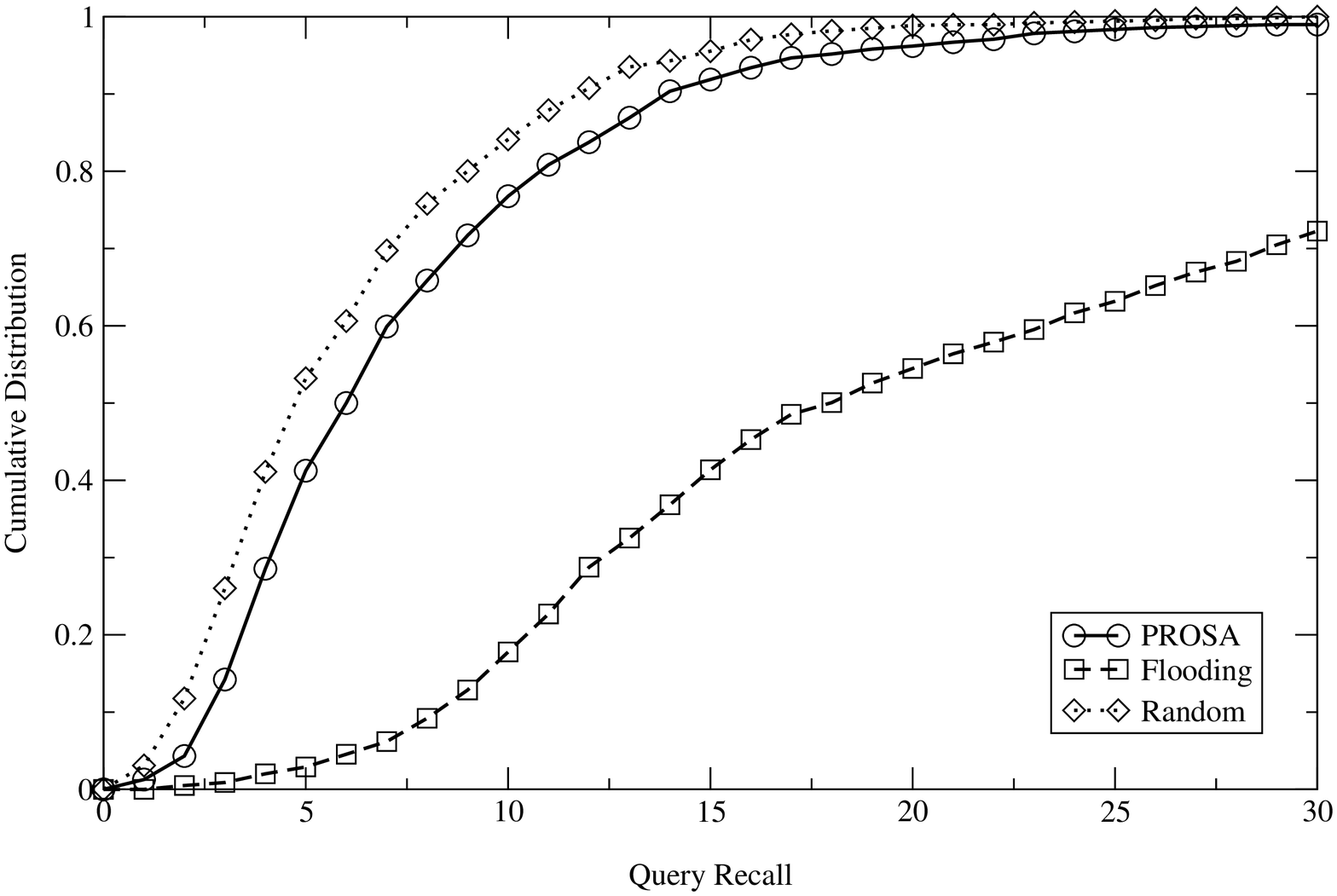}
    \label{fig:comm}
  }
  \hspace{0.1cm}
  \subfigure[Average query deepness]{
    \includegraphics[scale=0.22]{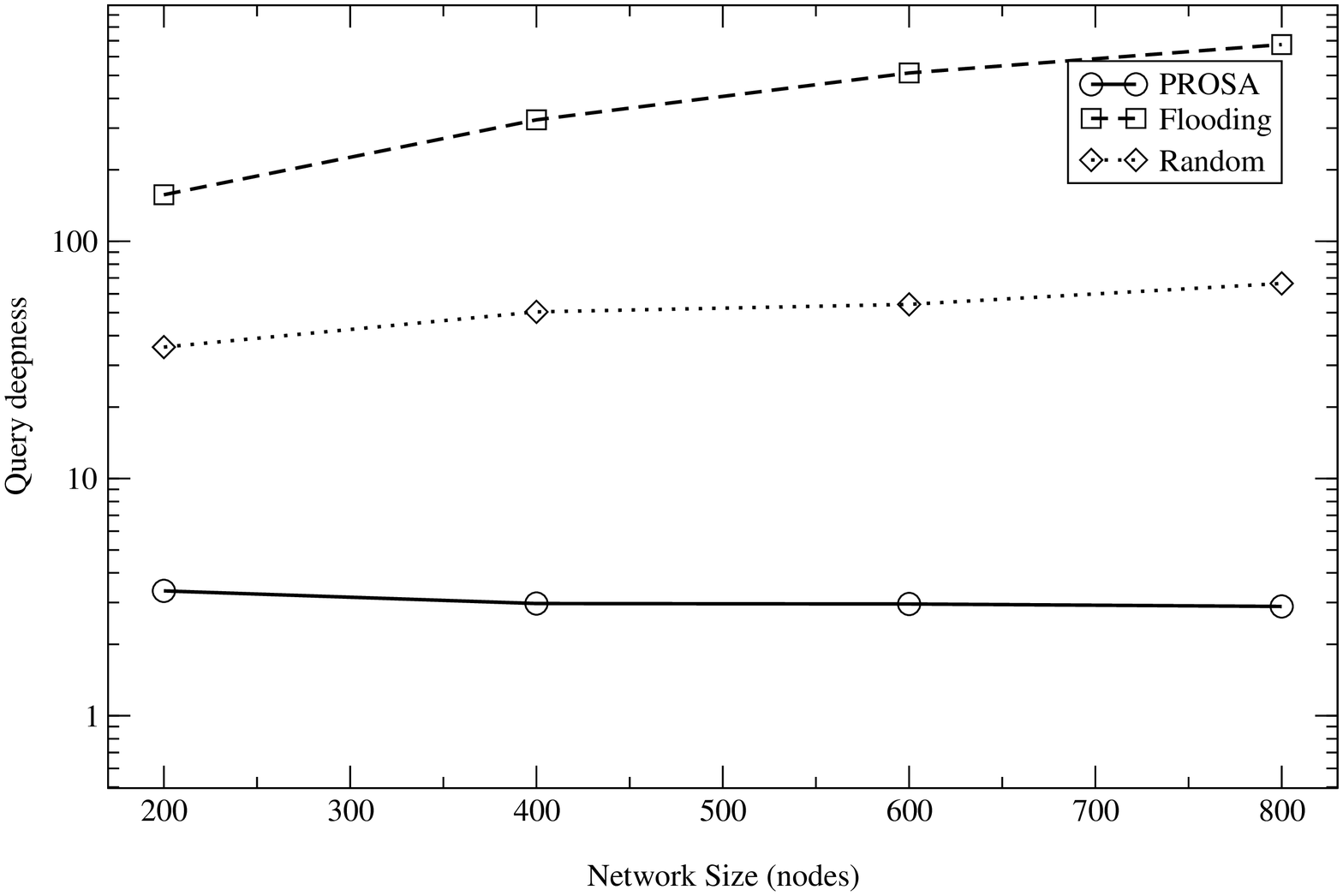}
    \label{fig:queryapl}
  }
  \caption{}
\end{figure}

Either if it is an important parameter for a resource searching and
retireving strategy, the number of retrieved documents is not the best
measure of how much documents a searching algorithm is able to
retrieve. Since not all queries match the same number of documents, it
is better to measure the percentage of retrieved documents over all
matching documents.  A valuable measure is the so--called ``recall'',
i.e. the percentage of distinct retrieved documents over the total
amount of distinct existing documents that match a query.  In figure
\ref{fig:recall} we show the recall distribution for \thealg, flooding
and random walk when each node performs 15 queries on average.

The best performance is obtained, once again, flooding the network:
about 60\% of queries have a recall of 100\%, and about 80\% of
queries have a recall of 50\%. Searching by flooding could not
return all documents because \thealg is a directed graph, and
unconnected components could still exist.  Also \thealg has high
recall: about 20\% of queries obtain all matching documents, while
45\% of queries are answered with one half of the total amount of
matching documents.  Random walk is the worst case: about 80\% of
queries has a recall of less than 50\% and only 8\% of queries obtain
all matching documents.

Recall measured as the simple percentage of retrieved document
over the total amount of matching documents does not take into account
the fact that in \thealg queries are requested to retrieve $n_r$
documents and no more. This fact could practically influence the
recall measure for \thealg networks, since queries are no more
forwarded if a sufficient number of documents has been retrieved. On
the other hand, it is important to analyse the recall in the case of
``rare'' queries. Note that we consider a query as beeing ``rare''
when the total number of matching documents is lower than the
number of requested documents; similarly a query is considered
``common'' if it matches more than $n_r$\footnote{Reported
results are relative to $n_r = 10$}.

Figure \ref{fig:rare} shows the cumulative normalised distribution of
recall for rare queries, while figure \ref{fig:comm} reports the
cumulative distribution for common queries.

Results reported in figure \ref{fig:rare} are really interesting:
\thealg answers 35\% of rare queries by retrieving all matching
documents, while 75\% of queries retrieve at least 50\% of the total
amount of matching documents; less than 10\% of queries obtain less
than 30\% of matching documents. Performance of a random walk is worse
than that obtained by \thealg: only 20\% of queries obtain all
matching documents, while more than 30\% of them obtain less than 30\%
of matching results.

The situation is slightly different for common queries. As reported in
figure \ref{fig:comm}, \thealg is able to retrieve at least 10
documents for 20\% of issued queries and, in every case, at least one
document is found for 99\% of queries, and at least 3 documents for
85\% of queries. We think that this behaviour is also affected by the
chosen value of $n_r$.

In order to better understand benefits of using \thealg, it is
interesting to look also at other measures that could clarify some
\thealg characteristics.  For instance, recall results are of poor
relevance without a measure of how fast answers are obtained.  A
feasible measure of speed could be the average query deepness, defined
as the average number of ``levels'' a query is forwarded far away from
the source node.

In figure \ref{fig:queryapl} we show average deepness of successfull
queries for \thealg, flooding and random walk on the same \thealg
network for different number of peers.

Query deepness for \thealg is around 3 and is not heavily affected
from the network size, while that of flooding and random walk is much
higher (from 30 to 60 and from 120 to 600, respectively).  Better
results obtained by \thealg cannot be simply explained by network
clustering coefficient, since all simulation are performed on the same
network. We suppose that it is mainly due to the searching algorithm
implemented by \thealg itself: it is able to find a convenient and
efficient route to forward queries along, avoiding a large number
of forwards to non--relevant nodes.

\section{Energetical Considerations}
\label{sect:energy}

An important parameter to take in account in order to quantify the
efficiency of a searching strategy is the ``energy'' needed to forward
and answer each query. In a theoretical model it is probably of no
great importance how much power is needed in order to answer a
query. But for real systems this is a crucial parameter. One of the
main issues with unstructured P2P networks such as Gnutella
\cite{Hybrid} is that queries waste a lot of bandwith, since a large
fraction of the network is flooded and a great amount of nodes are
involved in answering each query.  It is possible to roughly define
the average ``energy'' required for each query using equation
\ref{eq:energy}, where $N_q$ is the number of nodes to which the query
has been forwarded and $L_q$ is the number of links used during query
routing. $b$ and $c$ are dimensional scaling factors.

\begin{equation}
  E_q = b\cdot L_q + c\cdot N_q 
  \label{eq:energy}
\end{equation}

The definition given here for query energy is quite simple: it takes
into account the required bandwith, represented by the factor $b\cdot
L_q$, and the computational power needed by nodes in order to process
queries, represented by $c\cdot N_q$.

To estimate the amount of energy required to answer queries, we could
look at the average number of nodes and the average number of links
involved in each query.  Note that $N_q$ and $L_q$ are usually
different, since a node can be reached using many paths: either if it
processes the query only once \footnote{requests with the same query
id are ignored}, the bandwith wasted to forward the query to it cannot
be saved.

\begin{figure}[!htbp]
  \centering
  \subfigure[Average number of visted nodes per query]{
    \includegraphics[scale=0.22]{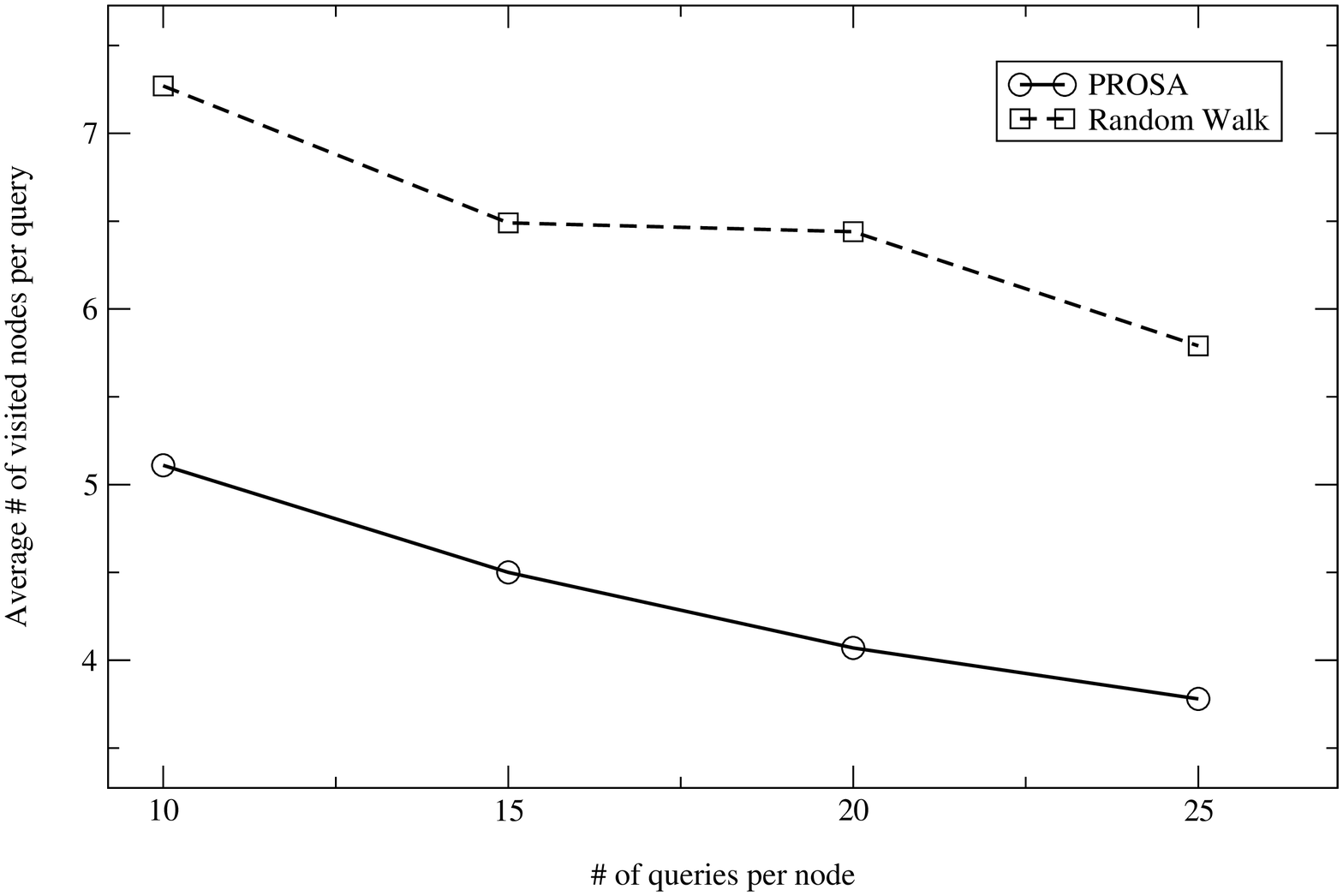}
    \label{fig:nodes}
  }
  \hspace{0.1cm}
  \subfigure[Average number of used links per query]{
    \includegraphics[scale=0.22]{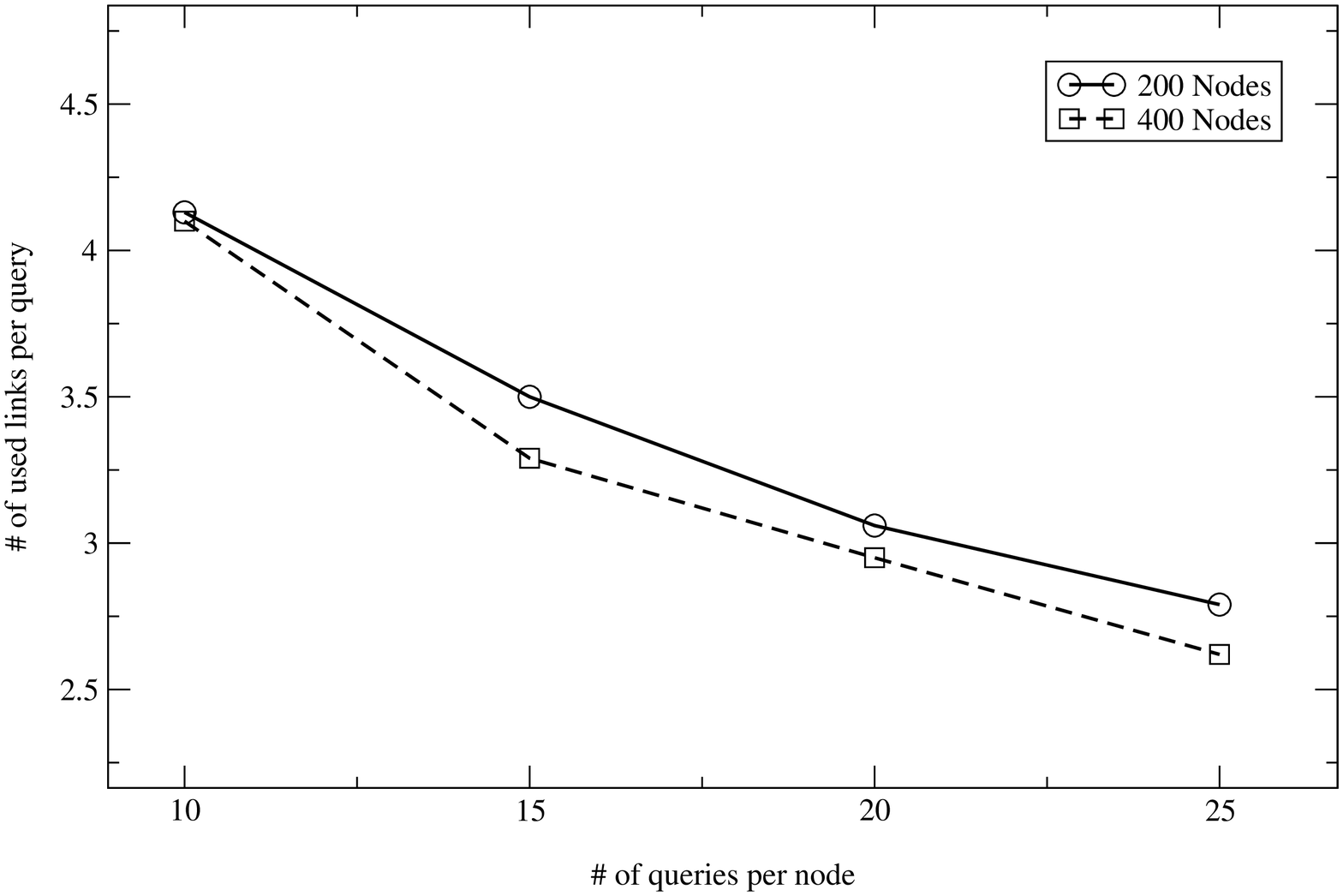}
    \label{fig:avglinks}
  }
  \caption{}
\end{figure}

Figure \ref{fig:nodes} and \ref{fig:avglinks} show, respectively.  the
average number of nodes involved and the average number of links used
by successful queries, both for \thealg and a simple random walk
search.

Since random walk uses a higer number of nodes and a higher number of
links in order to answer the same queries, it is clear that \thealg
requires less energy. On the other hand, since \thealg is able to
retrieve more matching documents than a random walk (as shown in
section \ref{subsect:numdocs}), we can state that \thealg is really
efficient with respect to average ``energy'' required to answer
queries.

\section{Conclusions and Future Work}
\label{sect:future}

This work presented a formal description of \thealg, a
self--organising system for P2P resource sharing heavily inspired by
social networks. Simulations show that resource searching and
retrieving in \thealg is really efficient, because of the ability of
peers in making good local choices that result in fast and successful
global query routing. Interesting results are obtained for query
recall measured on rare documents: \thealg is able to route queries
for those documents directly to nodes that probably can successfully
answer them. Since \thealg results to be a small--world, all nodes are
reached in a few steps, avoiding to waste bandwith and processing
power.  Future works include further studying \thealg in order to
discover emerging structures, such as semantic groups and communities
of similar peers.

\bibliography{article}
\bibliographystyle{plain}

\end{document}